# Design and analysis of a HTS internally cooled cable for the Muon Collider target and capture solenoid magnets

L. Bottura(1), C. Accettura(1), A. Kolehmainen(1), J. Lorenzo Gomez(2), A. Portone(2), P. Testoni(2)

(1) CERN, Geneva, Switzerland
(2) Fusion for Energy (F4E), Barcelona, Spain

## Abstract

The Muon Collider is one of the options considered as the next step in High Energy Physics. It bears many challenges, last not least in superconducting magnet technology. The target and capture solenoid is one of them, a channel of approximately 18 m length consisting of co-axial solenoid magnets with a 1.2 m free bore and peak field on axis of 20 T. One of the main concerns come from the nuclear radiation environment that may influence the stable operation of the coil, as well as its material integrity. Energetic photons cause large radiation heat load, of the order of several kW in the cold mass, and deposit a considerable dose, several tens of MGy. Neutrons cause material damage, at the level of $10^{-3}$ DPA. These values are at the present limit of superconducting coil technology. We describe here the conceptual design of the target and capture solenoid, focusing on the HTS cable design, which is largely inspired by the VIPER concept developed at MIT. We show how to address margin and protection, cooling and mechanics specific to the HTS cable selected.

## Introduction

The 2021 Update of the European Strategy for Particle Physics has identified five high priority R&D topics to be addressed towards the next step in High Energy Physics [1]. One of the topics identified [2] is the conceptual design of a Muon Collider (MC), a machine that could explore physics at the energy frontier. A MC can offer collisions of point-like particles at very high energies, since muons can be accelerated in a ring without the severe limitation from synchrotron radiation experienced by electrons. For center-of-mass energies in excess of about 3 TeV, a MC can provide the most compact and power efficient route towards a high luminosity collider at the energy frontier. However, the need for high luminosity faces technical challenges arising from the short muon lifetime at rest (2.2 µs) and the difficulty of producing bunched beams of muons with small emittance. Addressing these challenges requires a collaborative effort [3] towards the development of innovative concepts, especially in the field of superconducting magnets.

Among the several magnet challenges identified [4] one of the most demanding, and the focus of this paper, are the solenoids that host the target and capture channel, where the muon beam is produced. Muons result from the decay of positive and negative pions that are generated by the collision of a short, high intensity proton pulse with a solid target such as a carbon bar. The pion production target is inserted in a steady-state, high field solenoid, whose function is to capture the pions of both charges and guide them into a decay channel, where the muons are created. The magnetic field profile along the axis of the channel needs to have a specific shape, with peak field of 20 T on the target, and a decay to approximately 1.5 T at the exit of the channel, over a total length of approximately 18 m. The characteristic length of the field change is about 2.5 m, i.e. much larger than the gyration radius of the muons in the field so that the beam expands adiabatically in the channel.

The exceptional field level is the first, but not only challenge for the target solenoid. The interaction of the proton beam with the target produces a considerable amount of radiation, which needs heavy shielding to avoid heating and damaging the materials of the superconducting coils of the target solenoid. Such shielding, typically consisting of a combination of a heavy metal like tungsten and moderators like water, requires space, making the bore of the target solenoid large [5]. Nuclear calculations show that in order to host the target and sufficient shield, a superconducting solenoid would need a free bore of at least 1.2 m [6]. The large bore dimension corresponds to high stored magnetic energy, which in turn affects magnet protection, and electromagnetic forces.

In fact, field and bore are comparable to those of solenoids for fusion application. This is why for the design of the target and capture solenoids of the Muon Collider we have taken inspiration from the work performed for the central solenoid of fusion reactors such as ITER [7], and recent work towards compact machines such as ARC [8]. Furthermore, given the advances in HTS industrialization and magnet technology, we have selected a full-HTS solution as baseline hypothesis for our study. We recall here that the target solenoid of a Muon Collider operates in principle in steady state, which is a substantial difference with respect to fusion requirements of pulsed operation. Also, in fusion machines it is usually the coils close to the plasma that are most subjected to radiation, e.g. the toroidal magnet of a tokamak, and not the solenoid as in our case.

Besides the field reach, levels of 20 T are at the upper limit of performance for small bore $Nb_3Sn$, and arguably out of reach for LTS with the bore dimension required, the choice of HTS gives the possibility to set an operating point at a temperature higher than liquid helium. This brings the benefit of increased cryogenic efficiency, reduced wall-plug power consumption, and reduced helium inventory. We have taken for this study an operating temperature in the range of 20 K as representative of suitable cooling conditions.

In this paper we will describe briefly the electromagnetic and mechanical design of the target and capture solenoids. We will focus on the most challenging among them, i.e. the target solenoid, and select a suitable HTS cable geometry. We will then discuss force-flow cooling of the solenoid with gaseous helium at approximately 20 K, give estimates for the operating margin, report on quench detection and protection analyses, and complete our study with considerations on the internal mechanics of the cable. Note that the intention of the study reported here is to show that an all-HTS solenoid of this class is feasible, but the selected geometry and materials are not necessarily optimal. Further improvement of the magnetic configuration, of the mechanical reinforcement and support system, and of the cable operating point are possible, but beyond the scope of this initial work.

**Target solenoid magnetic design**

Previous work on the design of the target solenoid, performed within the scope of the US-DOE Muon Accelerator Program (MAP) [9], produced a design based on an LTS *outsert*, generating 14 T in a bore of 2 m diameter, and a resistive *insert*, generating the remainder field of 6 T, for a total of 20 T on the target [10]. The coils of the MAP design are shown schematically in Fig. 1, including a view of the decay and capture channel. The stored magnetic energy of this system is about 3 GJ, the estimated mass is about 200 tons, and the estimated power consumption is about 12 MW, including both the resistive insert and the cryogenic plant.

An alternative to the study performed by MAP is to assume an all-superconducting solenoid, using HTS operated at 20 K. With this choice it is in principle possible to reduce the radial build because of two reasons. Firstly, the resistive insert is no longer required, which also reduces considerably the power consumption. Secondly, operation at higher cryogenic temperature has higher thermodynamic efficiency than operation in liquid helium, a factor four to five if we compare 20 K to 4.2 K. This can be used to reduce the thickness of the nuclear shield, accepting higher heat loads in the cold mass, still resulting in lower total power consumption, and smaller magnet bore. We note at this point that the main concern is the photon fluence, yielding a heat load that is in the range of kW, and a radiation dose in the insulation system in the range of tens of MGy for operation over a reference period of ten years. The energetic particles generated at the target and their decay products, such as neutrons, are less of a concern, as the cumulated DPA remains below $10^{-3}$ for ten years of operation, a range in which the superconductor is not expected to degrade significantly [6].

Based on these considerations we have designed a coil system operating at 20 K with shielding thickness reduced by nearly a factor two compared to the solution of US-MAP, corresponding to a target solenoid bore of 1.2 m and total heat load in the cold mass of 4.1 kW [6]. To remove such cold heat load, a cryoplant operating at 20 K at its cold end would require approximately 300 kW of electric power at the warm end.

The coils layout is optimized to produce a reference (stationary) axial field along the z-axis ($B_z$) in accordance with the following expression [11]:

$$B_z = \frac{B_i B_f L_t^3}{B_i z^2 (3L_t - 2z) + B_f (L_t - z)^2 (2z + L_t)}, \text{ for } 0 < z < 15 \text{(m)} \tag{1}$$

where z is the coordinate measured from the center of the target, $B_i$ is the peak field, at the target location, $B_f$ is the field at the exit of the channel, and $L_t$ is a total length of the field decay along the solenoid axis. The reference design point has the following parameters: $B_i$=20 T, $B_f$=1.5 T, $L_t$=15 m. The field shape before the target, i.e. for z < 0, is not relevant, and the field is constant after the total decay length, i.e. for z > $L_t$, equal to $B_f$.

The coils layout in Tab. 1 achieves the field profile specified while minimizing the magnetic energy of the system. The system consists of three sections, respectively made of seven, eight and eight solenoid coils. The subdivision in three sections is mandatory for assembly reasons (i.e. maximum weight and handling length), as the whole set of 23 solenoids exceeds 18 m in length and, just the coil winding, weighs over 110 tons. For the calculation of the winding pack dimensions, it is assumed that the winding pack consists of stacked double pancakes. The conductor geometry and dimensions are described later. For the design of the winding pack we take a conductor turn-to-turn insulation of 1 mm, and a ground insulation of 20 mm. The three sections are further separated by a gap of 300 mm, needed for the mechanical structure and other connections. Four of the 23 solenoid coils, next to the gaps, are axially shorter, i.e. made of five double pancakes, the remaining 19 being made of 10 double pancakes. A total HTS conductor length of 8.75 km is required to build the whole magnet system. The conductor unit length in the longest double pancake is 150 m, including leads, which is comfortably shorter than the typical length of HTS tape that can be procured industrially.

Table 1. Geometry of the coils and operating current for the all-HTS solenoid. The winding pack of each solenoid module is centered at Rc/Zc and has dimensions dR/dZ. The winding pack is described by the number of turns and pancakes. The conductor current $I_{conductor}$ is intended as nominal nominal operating current.

| Coil | Rc (m) | Zc (m) | dR (m) | dZ (m) | Turns (-) | Pancakes (-) | $I_{conductor}$ (A) |
|---|---|---|---|---|---|---|---|
| C1 | 0.849 | -0.185 | 0.498 | 0.83 | 12 | 20 | 58905 |
| C2 | 0.87 | 0.665 | 0.54 | 0.83 | 13 | 20 | 60710 |
| C3 | 0.87 | 1.515 | 0.54 | 0.83 | 13 | 20 | 60392 |
| C4 | 0.808 | 2.365 | 0.415 | 0.83 | 10 | 20 | 51654 |
| C5 | 0.766 | 3.215 | 0.332 | 0.83 | 8 | 20 | 47469 |
| C6 | 0.704 | 4.065 | 0.208 | 0.83 | 5 | 20 | 46504 |
| C7 | 0.745 | 4.708 | 0.291 | 0.415 | 7 | 10 | 46293 |
| C8 | 0.704 | 5.423 | 0.208 | 0.415 | 5 | 10 | 53168 |
| C9 | 0.662 | 6.065 | 0.125 | 0.83 | 3 | 20 | 43280 |
| C10 | 0.662 | 6.915 | 0.125 | 0.83 | 3 | 20 | 42146 |
| C11 | 0.642 | 7.765 | 0.083 | 0.83 | 2 | 20 | 49452 |
| C12 | 0.642 | 8.615 | 0.083 | 0.83 | 2 | 20 | 44183 |
| C13 | 0.642 | 9.465 | 0.083 | 0.83 | 2 | 20 | 39567 |
| C14 | 0.642 | 10.315 | 0.083 | 0.83 | 2 | 20 | 32713 |
| C15 | 0.642 | 10.958 | 0.083 | 0.415 | 2 | 10 | 46717 |
| C16 | 0.642 | 11.673 | 0.083 | 0.415 | 2 | 10 | 45905 |
| C17 | 0.621 | 12.315 | 0.042 | 0.83 | 1 | 20 | 52310 |
| C18 | 0.621 | 13.165 | 0.042 | 0.83 | 1 | 20 | 56056 |
| C19 | 0.621 | 14.015 | 0.042 | 0.83 | 1 | 20 | 51602 |
| C20 | 0.621 | 14.865 | 0.042 | 0.83 | 1 | 20 | 51376 |
| C21 | 0.621 | 15.715 | 0.042 | 0.83 | 1 | 20 | 50471 |
| C22 | 0.621 | 16.565 | 0.042 | 0.83 | 1 | 20 | 52861 |
| C23 | 0.621 | 17.415 | 0.042 | 0.83 | 1 | 20 | 57438 |

The result of this design exercise is contrasted in Fig. 1 to the result of the MAP design, to identical scale. The schematic coil layout of the HTS solenoid is shown in Fig. 2, with a detail of the winding pack reported in the inset, as well as the conductor (described later).

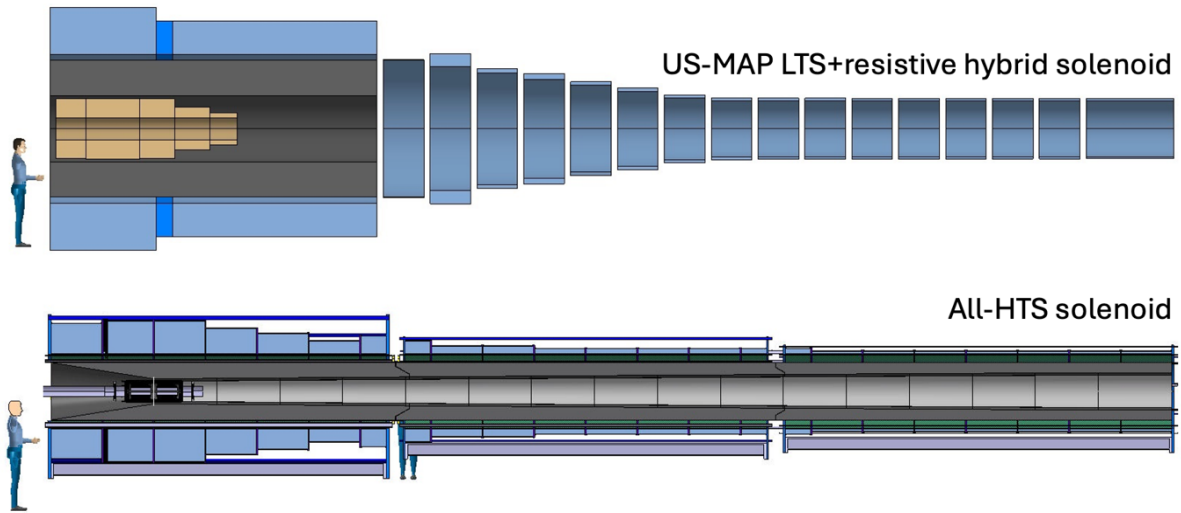

Figure 1. Comparison (to scale) of the solenoid coils of the target, decay and capture channel of a Muon Collider, as produced by the MAP study (top) [10] and resulting from the optimization of an all-HTS solution (bottom).

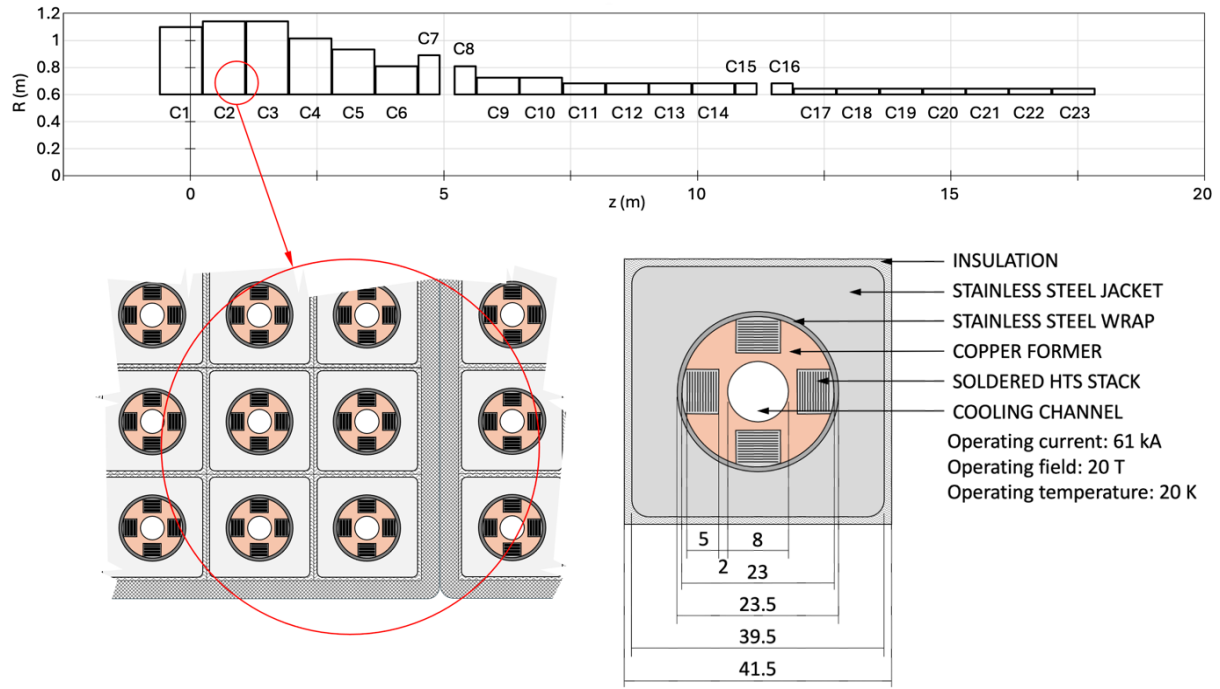

Figure 2. Schematic winding pack cross sections of the all-HTS coils in the target, decay and capture channel of the Muon Collider (top). The inset (bottom-left) reports a detail of the winding pack, showing the conductor with its turn/pancake insulation and ground insulation. We also report here the conductor cross-section (bottom-right, described later).

The field profile generated along the axis of the beam target decay and capture channel is shown in Fig. 3. As we see there, the field is very close to the reference field from beam physics, Eq. (1), and the profile generated by MAP. In the region where $B_z$>10 T, i.e. 0<z(m)<4 the field matching is within 0.5% of the target profile of Eq. (1). The shorter solenoid coils next to the inter-section gaps compensate for the missing ampere-turns, and achieve a maximum deviation of less than 4 % from the target profile around the gaps. On-going beam design studies have found this profile acceptable. The peak field on axis ($B_z$=20 T) translates in a peak field of 20.85 T on the most loaded cable of coil C2, which provides the dimensioning performance parameters for the conductor from the superconductor and structural standpoint (see sections below). The nominal operation currents for the various coils range from 33 kA (in coil C14) to about 61 kA (coil C2). The coils are intended as independently powered in this first design exercise, though we are aware that this leads to an excessive number of leads. Work is in progress to simplify this scheme. Further details on the coil geometry, currents, stored energy, and field quality are reported in the synoptic view of Fig. 4.

The striking difference in dimensions, clear from the visual comparison in Fig. 1, is reflected in a much smaller stored energy, 1 GJ for the HTS solution, lighter cold mass, about 110 tons, and reduced electrical consumption, estimated in the range of 1 MW for the cryogenics. We are aware that no such solenoid exists yet. Still, provided the technology can be demonstrated, this is a very clear motivation to pursue further the design, entering in the details discussed in the next sections.

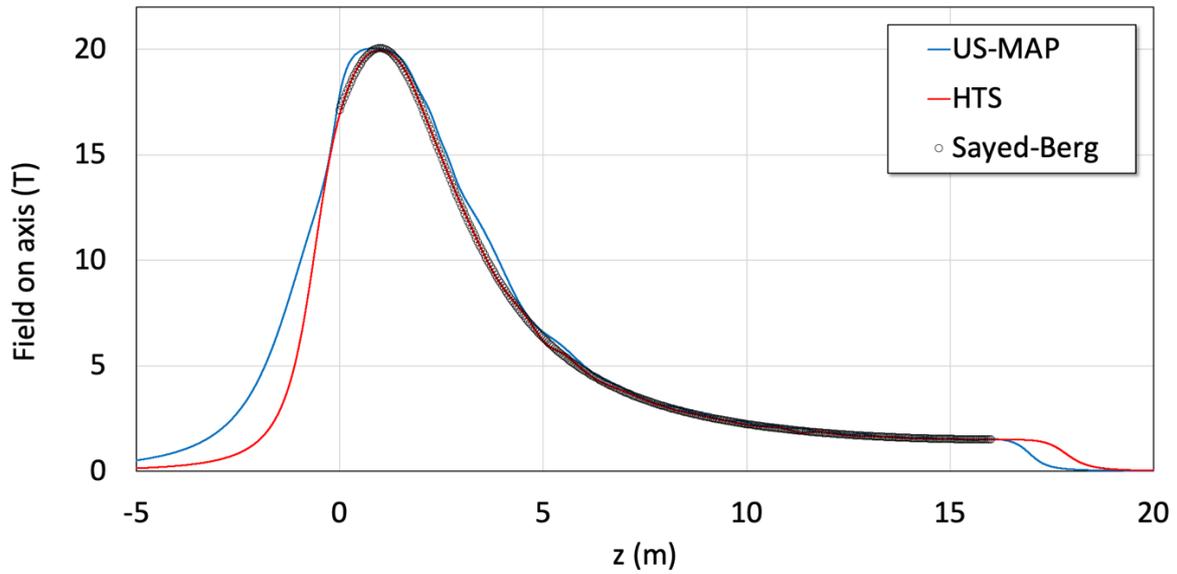

Figure 3. Comparison of field profile along the solenoid axis derived from Sayed-Berg [11] Eq. (1) (square symbols) as obtained for the MAP design [10] (blue line) and the all-HTS described here (red line).

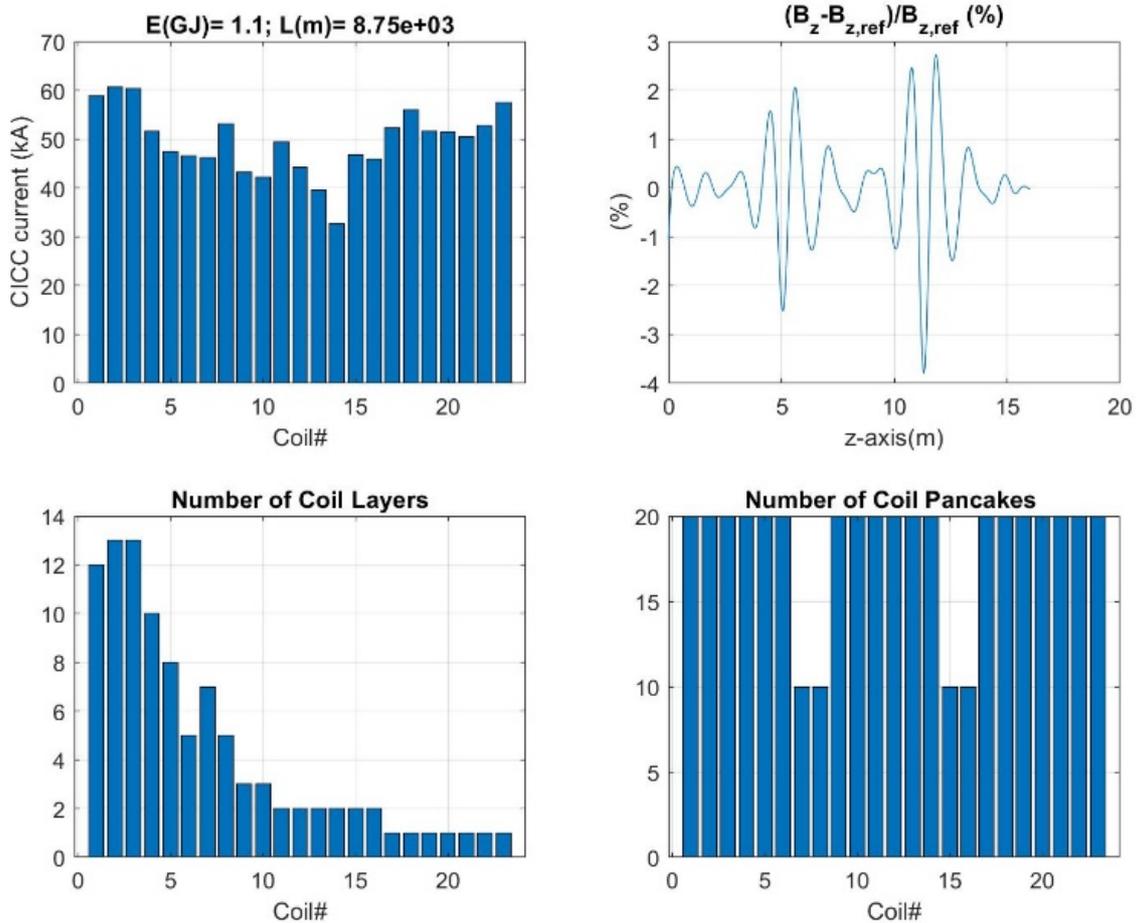

Figure 4: Top-left: conductors current, total magnetic energy (E) and total conductor length (L). Top-right: normalized field deviation from reference profile. Bottom-left: number of layers per coil. Bottom-right: number of single pancakes per coil.

**HTS cable layout and expected performance**

The cable configuration selected for the target solenoid is an adapted version of the VIPER cable developed by MIT [12], similar to the cable independently developed at ENEA [13]. The schematic layout is shown in Fig. 2, and the main geometry parameters are reported in Tab. 2. We have chosen a copper profile with a cooling hole of 8 mm diameter and 6 mm wide grooves. Each of the four grooves hosts a stack of REBCO tapes tightly packed and solder-impregnated in place during coil manufacturing. The depth of the groove is 5 mm, which is also the approximate height of the stack. Assuming a tape thickness of 60 μm, a typical value for present REBCO production, each stack would consist of 80 tapes. Tapes of different width and thickness could be used, which is inessential for this analysis provided the material fractions are maintained. Indeed, different stack dimensions and number of stacks may be convenient from the point of view of mechanics, though this matter is not explored in detail here. The grooves in the copper profile are twisted, so the stacks are transposed, though the single tapes are not. The target solenoid is a DC magnet, hence in principle twisting may not be mandatory for AC loss reduction [14], although ramping is still required for operation. The advantage of twisting in this case is to make the cable similarly flexible in all directions, because each tape sees on average the same deformation when wound in a pancake, which makes manufacturing easier. The assembly of copper profile and tape stacks is held in place by a stainless-steel wrap, applied with spacing, so to leave paths for the vacuum-pressure impregnation with soft solder. As proposed by MIT, the cable is first wound, and then impregnated once the tapes have taken their final position. An outer steel jacket is added to support the large hoop and compressive stresses generated by the electromagnetic forces. We will expand on the mechanical analysis and sizing of the jacket later. Because the helium cooling is contained by the copper profile, which guarantees tightness, the stainless-steel jacket does not need to be qualified for helium tightness, and can be conveniently manufactured by longitudinal welding of long extruded profiles.

Table 2. Composition and main geometric parameters of the cable.

| | | |
|---|---|---|
| HTS tape thickness | (μm) | 62 |
| HTS tape width | (mm) | 6 |
| HTS tape substrate (Hastelloy) thickness | (μm) | 40 |
| HTS tape copper thickness | (μm) | 20 |
| HTS tape YBCO thickness | (μm) | 2 |
| HTS stack width | (mm) | 6 |
| HTS stack thickness | (mm) | 5 |
| Number of HTS tapes | (-) | 80 |
| Number of HTS stacks | (-) | 4 |
| Copper profile diameter | (mm) | 23 |
| Cooling hole diameter | (mm) | 8 |
| Wrap thickness (no overlap) | (mm) | 0.25 |
| Jacket outer dimension | (mm) | 39.5 |
| Turn/pancake insulation | (mm) | 1 |
| Insulated conductor dimension | (mm) | 41.5 |

The critical properties of YBCO, temperature $T_{irr}$, field $B_{irr}$ and current density $J_C$, were obtained using the simple scaling law outlined in Tab. 3. The angular dependence was neglected in the analysis, also because the tapes twisted around the copper profile are periodically exposed to all field orientations. The parameters of Tab. 4 were used for the calculation, matching well field and temperature behavior of commercial HTS tapes in perpendicular field. With this choice of parameters, the critical engineering current density $J_E$ in the tape at 20 T and 20 K is 933 A/mm$^2$, and the corresponding conductor critical current

$I_C$ is 110 kA. We have set nominal operating conditions for the cable to 61 kA, 21 T and 20 K. This is not necessarily the optimum point, but is derived in a first attempt to prove design feasibility (i.e. 20 K operation at 20 T field) and reduce the inductance to keep dump voltage low (total 5 kV at the terminals, 2.5 kV to ground). The cable operates at about half the critical current and has a current sharing temperature of 29.7 K, i.e. a temperature margin of about 10 K. This is comfortable but required, as we will see later from the need to accept a large temperature increase when removing high heat flux at 20 K.

Table 3. Scaling law for the critical properties of YBCO in a field parallel to the crystallographic c-axis (normal) of the tape. Note that we indicate critical field $B_{irr}$ and temperature $T_{irr}$ as limits of irreversibility, as customary done for HTS materials.

| | |
|---|---|
| Critical irreversibility field | $B_{irr}(T) = B_{irr0}\left(1 - \dfrac{T}{T_{irr0}}\right)^\nu$ |
| Critical irreversibility temperature | $T_{irr}(B) = T_{irr0}\left(1 - \dfrac{B}{B_{irr0}}\right)^{\frac{1}{\nu}}$ |
| Critical current density | $J_C = \dfrac{C_0}{B} h(t) f_p(b)$ |
| Reduced field | $b = \dfrac{B}{B_{irr}(T)}$ |
| Reduced temperature | $t = \dfrac{T}{T_{irr0}}$ |
| Scaling of pinning force with reduced temperature | $h(t) = (1 - t^\nu)(1 - t^m)$ |
| | $f_p(b) = \dfrac{1}{f_{p,max}} b^p (1 - b)^q$ |
| Scaling of pinning force with reduced field | $f_{p,max} = \left(\dfrac{p}{p+q}\right)^p \left(1 - \left(\dfrac{p}{p+q}\right)\right)^q$ |

Table 4. Parameters used to compute the critical properties of YBCO. Note that the critical current density multiplier $C_0$ is defined referring solely to the superconducting (YBCO) cross section.

| | | |
|---|---|---|
| $C_0$ | (A T /mm²) | 2270000 |
| $B_{irr0}$ | (T) | 274.89 |
| $T_{irr0}$ | (K) | 91.3 |
| $m$ | (-) | 0.701 |
| $\nu$ | (-) | 0.332 |
| $p$ | (-) | 0.75 |
| $q$ | (-) | 5.7 |

## Cooling at 20 K

The potential of cooling at high cryogenic temperature is generally associated to the well-known increase of the Carnot efficiency as the temperature of the cold point of the heat engine is increased. Correspondingly, the Coefficient Of Performance (COP), the ratio of the heat removed from the cold point to the work necessary at room temperature, is expected to increase by a factor of about five between 4.2 K and 20 K. However, while this may be true in the case of the thermodynamic cycle of the refrigerant in the cryogenic plant, one has to consider that cooling also requires distribution, possibly at high mass-flow (because of reduced density in gas at high temperature) and increased pressure drop (because of high flow velocity). The work done to distribute the fluid, which is necessary to remove heat, appears as an equivalent heat load at the level of the cryogenic plant. This additional heat load affects the COP of the overall installation. Indeed, a high temperature cooling system with sub-optimal design of the distribution may yield a COP which is not much different from that of a system operating at a lower temperature, but better optimized distribution.

To quantify the above considerations, we consider the case of a single channel where a pump circulates helium, removing a quantity of heat $\dot{q}_{load}$, and re-cooled to the inlet temperature by a heat exchanger. The work done by the pump $\dot{q}_{pump}$ recirculating a mass-flow $\dot{m}$ in the channel, under a pressure drop $\Delta p$ is:

$$\dot{q}_{pump} \approx \frac{\dot{m}}{\langle\rho\rangle\eta_{pump}}\Delta p \qquad (1)$$

where we have used an average density $\langle\rho\rangle$ to take into account that the state of the coolant changes in the channel, and $\eta_{pump}$ is the isentropic efficiency of the pump. Equation (1) is the additional work that the refrigerator must remove at the cold end.

The COP of a refrigerator is commonly defined as the ratio of the heat removed at the cold end of the system, $\dot{q}_{cold}$, to the power required at the warm end of the refrigerator to run the machine, $W_{warm}$. A high efficiency corresponds to a large *COP*. In an ideal case the power seen at the cold end of the refrigerator is equal to the heat load in the system, i.e. $\dot{q}_{cold} = \dot{q}_{load}$ so that:

$$COP_{ideal} = \frac{\dot{q}_{load}}{W_{warm}} \qquad (2)$$

In the case of force-flow cooling the effective coefficient of performance of the system $COP_{real}$ needs to include the additional contribution originated by the pump work, i.e. $\dot{q}_{cold} = \dot{q}_{load} + \dot{q}_{pump}$. Simple algebra yields [15]:

$$COP_{real} = COP_{ideal}\frac{\dot{q}_{load}}{\dot{q}_{load}+\dot{q}_{pump}} \qquad (3)$$

As indicated in Eq. (3), a force flow system will always have a *COP* smaller (lower efficiency) than that of an ideal cryogenic system removing only the power $\dot{q}_{load}$. It is important to maintain the pump work small with respect to the heat removed, so that *COP*$_{real}$ remains close to *COP*$_{ideal}$.

Let us explore further the scaling of $\dot{q}_{pump}$. We can write simple approximations for the pressure drop and the mass-flow required, i.e.

$$\Delta p \approx \frac{2fL}{D_h}\frac{\dot{m}^2}{A^2\langle\rho\rangle} \qquad (4)$$

$$\dot{m} \approx \frac{\dot{q}_{cold}}{c_p \Delta T} \qquad (5)$$

Where *f* is the Fanning friction factor, *L* is the length of the channel, $D_h$ is the hydraulic diameter of the channel, *A* is the cross section of the channel, $c_p$ is the specific heat of the fluid at constant pressure and *ΔT* is the temperature increase from inlet to outlet of the channel. Above expressions are approximate, but appropriate for scaling analysis. We assume perfect gas state to compute the density dependence on temperature and pressure. This is an acceptable hypothesis for helium at 20 K, where the compressibility ratio only deviates by a few percent from the ideal value up to large pressures (maximum deviation of 7.5 % in the range of 1 bar to 30 bar). We also assume constant transport properties (*f* and $c_p$), and we arrive at the following interesting scaling relation, where we omit the scaling constant:

$$\dot{q}_{pump} \propto \left(\frac{\dot{q}_{cold}}{\Delta T}\right)^3 \left(\frac{\langle T \rangle}{\langle p \rangle}\right)^2 \qquad (6)$$

where we used average values of temperature <*T*> and pressure <*p*>. Equation (6) deserves some comments. The pump work grows rapidly, with the third power of the heat removed from the channel (the first term). Looking at relatively large heat loads, the only way to reduce the pump work is to accept larger temperature increase from inlet to outlet. The second notable dependence is on temperature (the second term). An increase in temperature causes a decrease in density which results in larger pressure drop for a given mass-flow. The dependence on temperature is also rather steep, with the second power. Recall that the COP of a refrigerator only increases with the inverse of the temperature of the cold source. If all other conditions were kept equal, Eqs. (3) and (6) imply that increasing the temperature of the cold end we would reach a point when the COP would decrease (i.e. lower efficiency) rather than increase (i.e. higher efficiency). Back to Eq. (6), we can see that a way to avoid this is to increase the pressure, maintaining the ratio <*T*>/<*p*> constant. High cryogenic temperature force-flow systems require high pressure.

We have put the above scaling in practice by taking typical values relevant to our study, assuming a channel length of 150 m, a heat load of 150 W, and channel diameter of 8 mm. The results of a scan of inlet temperature and outlet pressure, assuming a temperature increase of 3 K, are shown in Fig. 5. The legends indicate the expected scaling from Eq. (6), which are well verified also in the practical case once helium properties are considered. A quantitative evaluation can be seen in Fig. 6, which reports slices of the surface of the type reported in Fig. 5, either at constant inlet temperature (left) or constant outlet pressure (right). We see from the curves and values reported there that in order to maintain a pumping power *small* e.g. of the order of 10 %, of the removed heat, we need to accept temperature increase of the order of several degrees (e.g. 3 K or higher) and operate at high pressure (e.g. 20 bar or higher). We also see from Fig. 6 that decreasing the operating temperature has advantages, as expected. Adjustments of the operating point should be considered in order to achieve the optimal overall performance. This is part of a global optimization, an important step but outside of the scope of this work.

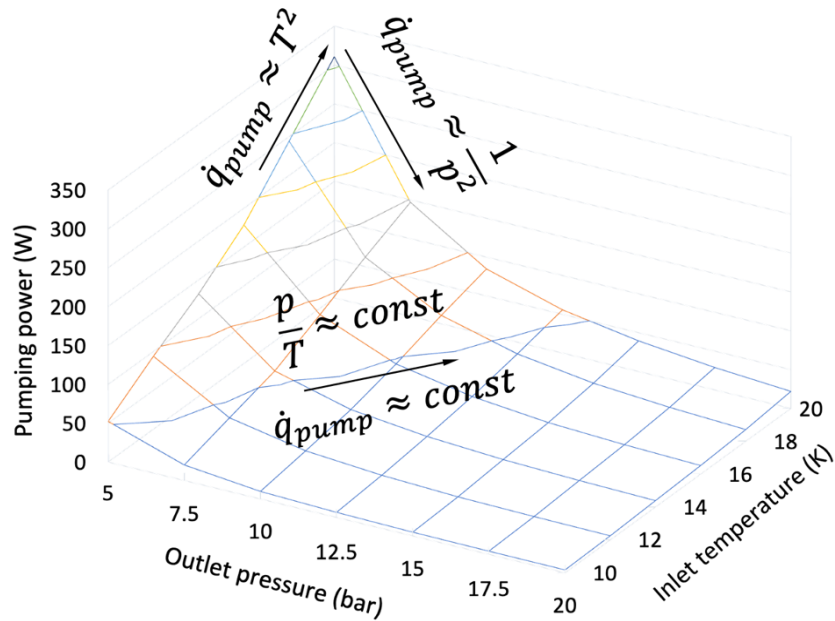

Figure 5. Pumping power as a function of inlet temperature and outlet pressure for helium cooling of a 150 m long channel of 8 mm diameter, heat load of 150 W and allowed temperature increase of 3 K.

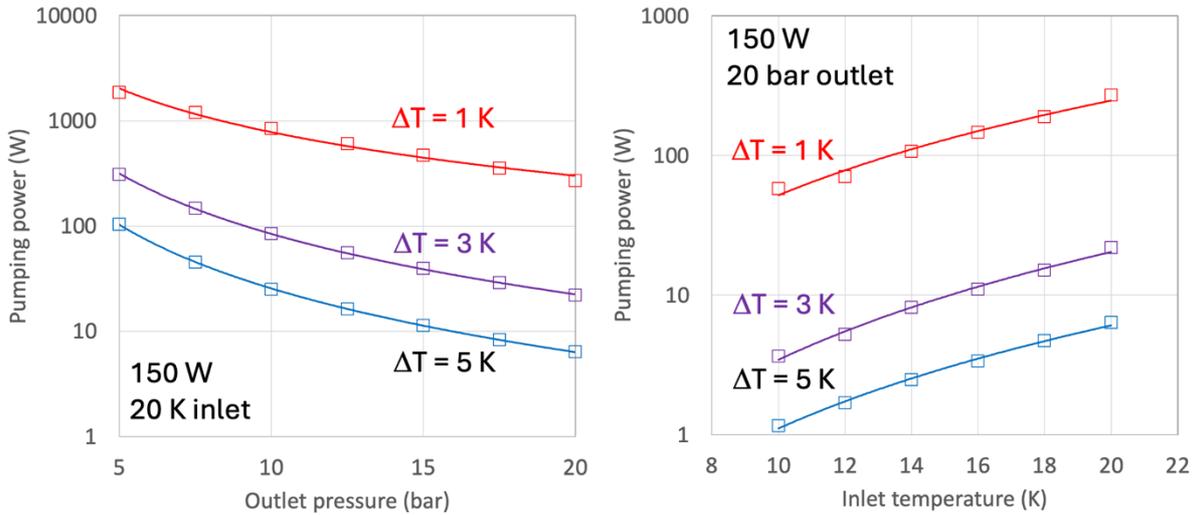

Figure 6. Pumping power as a function of outlet pressure at constant inlet temperature, and as a function of inlet temperature at constant outlet pressure, for helium cooling of a 150 m long channel of 8 mm diameter, heat load of 150 W and several values of allowed temperature increase.

We complete our analysis of cooling at 20 K with a practical calculation where we have taken the heat distribution that approximates energy deposition from radiation in the most heated portion of the target solenoid. We assume that the solenoid is assembled from double pancakes, with helium inlet and outlet at the outer radius. Simulations were performed using the code THEA [16]. The THEA model is relatively simple, giving the scoping nature of this analysis. It consists of a single thermal component and a single hydraulic component. The thermal component represents the round copper profile and the HTS tapes, and consists of three materials with identical temperature: the copper in the profile and in the HTS tapes, the Hastelloy in the HTS tapes, and the HTS material (YBCO). The hydraulic component corresponds to the helium flowing in the round cooling hole in the copper profile. Thermal and hydraulic components are coupled by heat convection at the wetted perimeter. Hydraulic diameter, wetted perimeter, and transport correlations are as from standard practice in round pipes. Note that we neglect the presence of the stainless-steel jacket for this analysis, as well as the counterflow heat exchange effect among the turns of the double pancake. The heat resistance among two adjacent turns is of the order of 1 K/W at 20 K, and the coupling across the turn insulation only has a marginal effect on the heat balance. Boundary conditions for the thermal components are adiabatic, while for the hydraulic component they are of prescribed pressure, temperature and massflow at inlet ("flow" condition in THEA) and pressure and temperature at outlet ("reservoir" condition in THEA). A rather coarse mesh of 200 linear elements was used, with implicit adaptive time stepping. The most salient parameters used for the simulation are reported in Tab. 5.

Table 5. Main parameters and options used for the cooling simulations with THEA.

| | | |
|---|---|---|
| Total HTS (YBCO) cross section | (mm$^2$) | 4.2 |
| Reference electric field for superconductor transition | (μV/m) | 10 |
| Reference n-power for superconductor transition | (-) | 30 |
| Total copper cross section | (mm$^2$) | 361 |
| Copper RRR | (-) | 20 |
| Total substrate (Hastelloy) cross section | (mm$^2$) | 77 |
| Helium cross section | (mm$^2$) | 50 |
| Hydraulic diameter | (mm) | 8 |
| Wetted perimeter | (mm) | 25 |
| Friction factor correlation | (-) | Blasius |
| Heat transfer correlation | (-) | Dittus-Boelter |

The main source of heating is of nuclear nature, i.e. the high energy particles and photons generated by the proton beam impinging on the target. The coil sees mainly photons (gamma rays) and neutrons. Most of the heating power is deposited by photons, while the main concern for neutrons is the damage on the superconductor [6]. We show the heating profile assumed in Fig. 7, developed along the length of a double pancake. We have selected for this analysis the double pancake that receives most radiation, in coil C2. From low values at the outer radius (inlet) the power increases towards the middle of the double pancake (inner radius) to decrease again when moving towards the outer radius in the second half of the double pancake. The total heat per double pancake is about 150 W, i.e. a considerable average heat load of 1 W/m. In our analysis we assume steady state heating conditions and ignore the time structure of the heat deposition. Other heating sources are negligible with respect to nuclear heat.

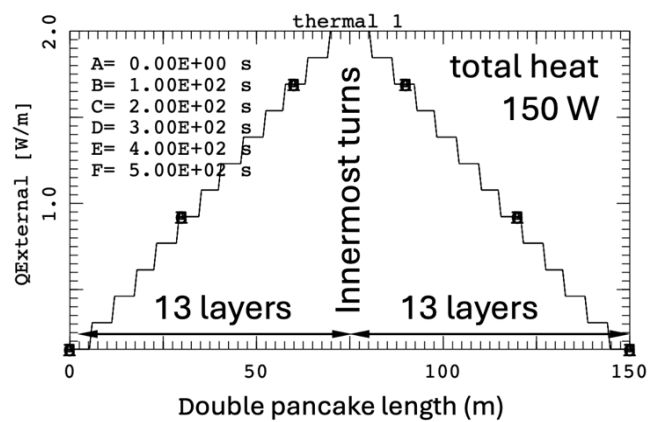

Figure 7. Profile of the heat deposition profile in the double pancake with most heat deposited.

The transient evolution of temperature in the channel is shown in Fig. 8, from the initial time when temperature is constant, to the steady state value after several residence times. The temperature rise is monotonic, maximum temperatures are reached in steady state. The mass-flow required to achieve a temperature increase of 3 K is about 8.7 g/s. Under this flow, and assuming an isentropic pump efficiency of 80 %, the heat load associated with the pumping power is 18 W, i.e. small with respect to the 150 W removed. Note that the inlet temperature selected is 18.5 K, which in case of a temperature increase of 3 K from inlet to outlet results in 20 K at the high field location, in the middle of the double pancake. The values obtained here, which consider the variable helium properties and transport correlations, are consistent and confirm the simplified scaling analysis discussed above. Based on this analysis we have set the nominal operating point for the cooling of the target solenoid at a inlet temperature of 18.5 K, temperature increase of 3 K, corresponding to a temperature of 20 K in the high field region, outlet pressure of 20 bar and mass-flow of 8.7 g/s per double pancake.

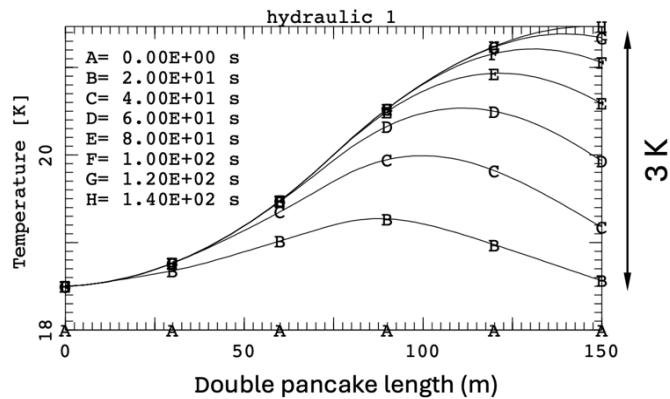

Figure 8. Transient temperature increase in the double pancake with most heat deposited, from initial conditions of constant temperature to steady state conditions after sufficiently long time (approximately 150 s).

**Stability study**

It is well known that HTS conductors have exceedingly large stability. We have quantified the value of the margin by performing calculations of the stability margin of the cable described earlier, and in the pancake with the highest field at inner radius. We define stability margin as the value of an external energy input over a specified length of cable, just sufficient to initiate a thermal runaway.

Simulations were performed using the code THEA [16] with a model of the same geometry as used for the cooling analysis, and the nominal cable performance discussed earlier, including the scaling of the critical current with field and temperature. The question rises on whether neglecting the stainless-steel jacket and insulation is appropriate. We can resolve it by evaluating the thermal diffusivity of the three main materials of the conductor, namely copper, stainless steel and glass-epoxy insulation, at 20 K. They are respectively 4.5 $10^{-3}$ m$^2$/s for copper, 1.7 $10^{-5}$ m$^2$/s for steel and 2.2 $10^{-6}$ m$^2$/s for glass-epoxy. Taking the radius of the copper core, the wall thickness of the steel jacket and the thickness of insulation as characteristic length scales, we can estimate the characteristic times of the temperature diffusion. We obtain a few ms for copper, about 1 s for stainless-steel and a few tens of s for glass-epoxy. This implies that on time scales relevant to thermal stability, a few ms to fractions of s, the temperature profile is flat inside the copper profile, as assumed in the model, the jacket temperature only changes over a short internal layer, with little influence to the heat balance, and the insulation is definitely not contributing.

The analysis of stability is of local nature (transients are relatively fast and there is little influence from outside boundaries) and requires fine discretization of the temperature profile to avoid meddling physical results with numerical error. For this reason, we have limited the analysis to a domain of 4 m length heated in its middle, of which only half of the domain (2 m) was meshed, imposing symmetry. Linear elements of 1 cm and a maximum time step of 10 ms were necessary to yield converged results. Conditions of steady current and constant field were considered in the domain analyzed. Boundary conditions for the thermal components were adiabatic at the symmetry axis, and constant temperature at the far end. Constant pressure and temperature reservoirs were assumed at both ends of the hydraulic component.

For the first parametric study we have set nominal operating conditions for current (61 kA) and temperature (20 K), varied the field, and computed the corresponding thermal stability margin for a fast perturbation, 1 ms. Each computed value represents a point in the coil located at a different length in the double pancake. The result of this scan is shown in Fig. 9. As expected, the stability margin for short (0.1 m) and relatively long (1 m) initial normal zone (INZ) is very high, of the order of several to several tens of $J/cm^3$. At these levels it is unlikely that an external heat input can ever quench the cable. The margin grows substantially at low field, where the temperature margin reaches values of 40 K and higher. Given this high margin, quench propagation speed is low, and the protection analysis is especially interesting for the low field region of the solenoid.

It is also interesting to note that the stability margin appears to be largely in excess of the enthalpy of the cable volume corresponding to the INZ, including or not the helium. This is due to two effects. The large copper profile contributes to strong longitudinal conduction, which spreads the heat beyond the INZ during the time of heating end the ensuing thermal transient, thus making the INZ effectively longer. Moreover, the transient tends to be slow, lasting fractions of s and longer. On this time scale the helium flow is sufficiently fast to remove heat through convection.

In Fig. 10 we have studied the influence of operating temperature on stability, considering the possibility that the coil is operating at nominal current, but the temperature may not be at the nominal value. Besides the trivial result that the stability margin decreases at increasing operating temperature, it is interesting to note that the decrease (or increase) of stability margin is not directly proportional to the decrease (or increase) of temperature margin. Indeed, increasing the operating temperature to 25 K, which corresponds to a temperature margin of about 5 K, i.e. half the nominal value, we obtain a stability margin of 2 $J/cm^3$, i.e. a reduction of only 20 % with respect to the value at nominal operation. Similarly, operating at 15 K, which corresponds an increase of the temperature margin by 50 %, only yields an increase of the stability margin by 15 %. We attribute the non-linear dependency to the fact that material properties are also changing considerably in this temperature range, in particular specific heat and conductivity.

In summary, the conductor proposed has very high stability at the operating point taken as reference, throughout the coil, and is tolerant to significant changes in the operating temperature. It is unlikely that the conductor can be quenched by external heat inputs. In fact, the exceptional stability at low field may hinder quench propagation and detection of normal zone, if initiated, which is the subject discussed in the next section.

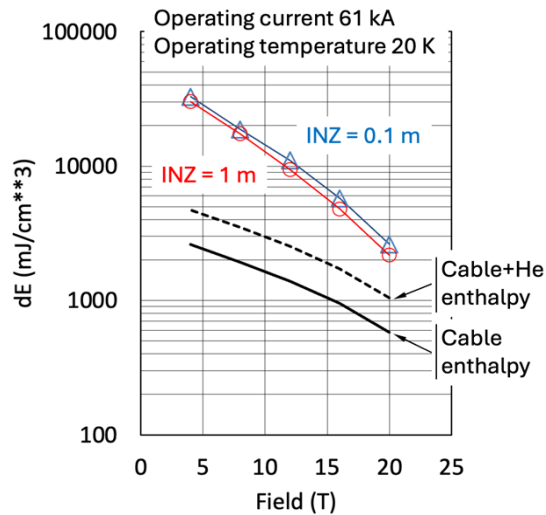

Figure 9. Stability margin of the target solenoid cable as a function of field at nominal conditions of current (61 kA) and temperature (20 K). Results for an INZ of 0.1 m (blue triangles) are compared to those for an INZ of 1 m (red circles).

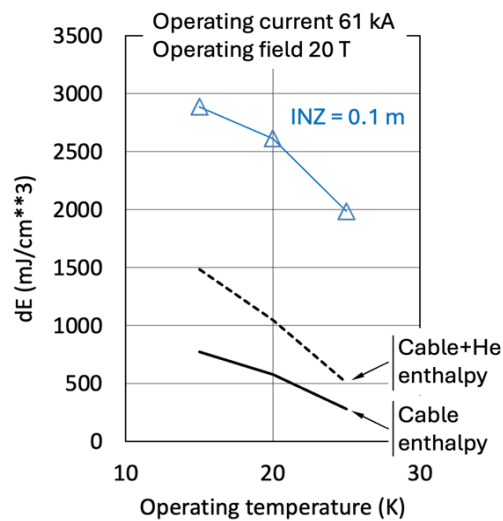

Figure 10. Stability margin of the target solenoid cable as a function of temperature at nominal conditions of current (61 kA) and field (20 T). Results for an INZ of 0.1 m (blue triangles).

**Quench and protection study**

We have anticipated that it will be difficult, if at all possible, to quench the conductor by heat deposition, i.e. stability is arguably not an issue for the HTS cable proposed here. Still, performance degradation associated with excessive stress and strain during manufacturing, thermal or electromagnetic cycles, or simply loss of nominal cryogenic operating conditions could result in a quench. It is hence important to study also whether a quench, initiated by reasons others than stability, can be detected and protected. For this scoping study we have considered a single pancake, operating at nominal conditions of peak field of 20 T and 20 K. To set the dump time constant $\tau_{dump}$ we have assumed a terminal voltage of 5 kV (2.5 kV to ground) and we have taken the coil with the largest stored energy, coil C2 in Tab. 1, which also corresponds to the coil with highest field and energy deposition. With a stored energy $E$ of about 300 MJ, a conductor operating current of $I_{op}$ 61 kA and a dump voltage $V_{dump}$ of 5 kV this leads to a minimum dump time constant of 2 s as can be obtained from the following relation that is valid when the external dump resistance is much larger than the quench resistance in the coil:

$$\tau_{dump} = \frac{2E}{I_{op}V_{dump}} \tag{7}$$

Simulations were performed using again the code THEA [16]. The conductor discretization is the same as used for cooling and stability, i.e. one thermal component and one hydraulic component, again neglecting the presence of the stainless-steel jacket and the thermal coupling among turns, which is a conservative assumption for quench. A total length of 150 m was analyzed, corresponding to the full double pancake. Adiabatic boundary conditions were applied to the thermal component, while for the hydraulic component they were reservoirs at constant pressure and temperature. The current was assumed to stay constant during the initial phase of quench initiation and propagation, and then dumped with a time constant of 2 s once the detection threshold was reached. Quench simulation requires careful choice of space and time integration, to reduce the influence of numerical diffusion on quench propagation that could corrupt the solution [17]. The initial mesh was refined in the location of the INZ, with 1.5 mm linear elements. Adaptive meshing at the quench front was used, with minimum mesh size of 1 mm. Adaptivity was also used for time integration, with implicit time stepping and an initial time step of 10 μs, increasing in time up to 100 ms.

We have initiated a quench by heating a length of 5 cm in the center of the pancake with an energy just above the stability margin and energy deposition over 1 ms. This can be taken as a quench representative of a very local degradation, as the INZ is close to the transverse size of the conductor, of the order of 4 cm. We have then studied the evolution of the normal zone and resistive voltage, triggering a coil dump on an external resistor as soon as the resistive voltage exceeded a detection threshold.

An example of the results of quench initiation and dump is shown in Fig. 11, where we have reported the evolution of the resistive voltage and cable current, of the normal zone and total propagation speed, and of the conductor and helium temperature at the hot spot, in the INZ. The normal zone is defined as the total length of cable where the temperature is above current sharing. A detection threshold of 100 mV was taken for the simulation. Note that the normal zone and propagation speed reported in Fig. 11 are the total value, i.e. two-sided, as the quench propagates up- and down-stream of the INZ. The front propagation is not necessarily the same in the two directions, as in principle the helium motion can contribute to the propagation velocity. In fact, in the range of conditions analyzed we found that helium motion has negligible influence on the normal zone propagation. The actual quench front speed is hence close to half the normal zone growth rate.

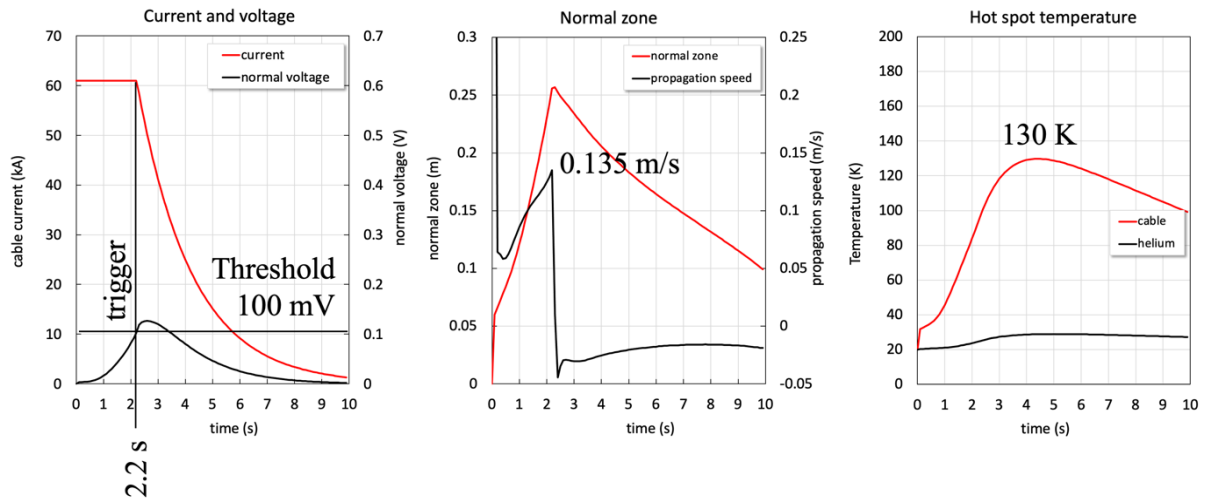

Figure 11. Example of evolution of a quench for operation at nominal current (61 kA) field (20 T) and temperature (20 K), assuming a quench initiated over 5 cm and detection threshold of 100 mV.

In the example of Fig. 11 the trigger is generated at a time of 2.2 s, once the resistive voltage reaches the threshold of 100 mV. The normal zone grows from an initial value of about 5 cm to about 25 cm after 2 s, a modest increase corresponding to an average front speed of 5 cm/s (about 10 cm/s two-sided). The maximum front speed peaks at about 7 cm/s (13.5 cm/s two-sided) at about 2 s. This low value is not surprising, it is well known that quench propagates slowly in HTS cables because of the high stability, as demonstrated in the previous analysis. Still, the combination of propagation and, especially, temperature increase generates enough resistive voltage for a reasonable detection. At the trigger time the conductor temperature is below 100 K, with enough headroom for a dump. In Fig. 11 we show that the hot spot only reaches a peak of 130 K, before being "washed away" by the helium flow.

Simulations of the type above have been repeated parametrically, taking the detection voltage as the free parameter. The results are shown in Fig. 12. We see there that the hot spot temperature can be maintained below 200 K, i.e. modest thermo-mechanical effects, for a detection voltage up to 200 mV. A range of 100 to 200 mV per each double pancake should be appropriate for a magnet of this type, which is operated in steady state.

A second study performed is to consider quench initiated at nominal current but in zones of lower field, or in points along the magnet loadline at both current and field below nominal. Sample results are shown in Tab. 6. We see from there that although quench propagation is very slow at low field, indicated by the longer detection time, it is still possible to protect the solenoid with the nominal dump and maintain a hot spot temperature below a reasonable value in the range of 170 K. In. fact, at low current the detection of a quench takes a very long time, nearly 10 s. But the corresponding Joule heating is also low, so that a dump yields a very safe hot spot temperature of 140 K.

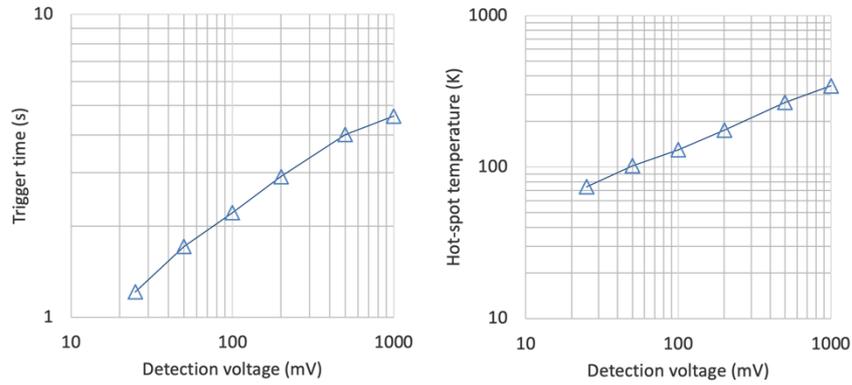

Figure 12. Parametric analysis of detection (trigger) time and hot spot temperature as a function of detection voltage threshold. The quench is initiated at nominal operating conditions of current (61 kA) field (20 T) and temperature (20 K), assuming a heated zone of 5 cm.

Table 6. Main results of quench simulation in nominal conditions at peak field, at low field (4 T) and during ramp at half current (30 kA) and field (9.84 T). Detection voltage of 100 mV assumed.

| Current (kA) | Field (T) | Detection time (s) | Hot spot temperature (K) |
|---|---|---|---|
| 61 | 20 | 2.2 | 130 |
| 61 | 4 | 2.8 | 172 |
| 30 | 9.84 | 14.8 | 140 |

## Mechanics

During steady state operation the Target Solenoid is predominantly affected by the hoop stresses reacting the large Lorentz forces applied on the cables, and by the axial compressive stresses due to the magnetic interaction among the different coils, as shown in Figure 13. The stresses are mainly supported by the conductor jacket, but also by the cables as their contribution to the overall stiffness is not negligible. For the proposed configuration of coil currents, no repulsive forces arise between coils.

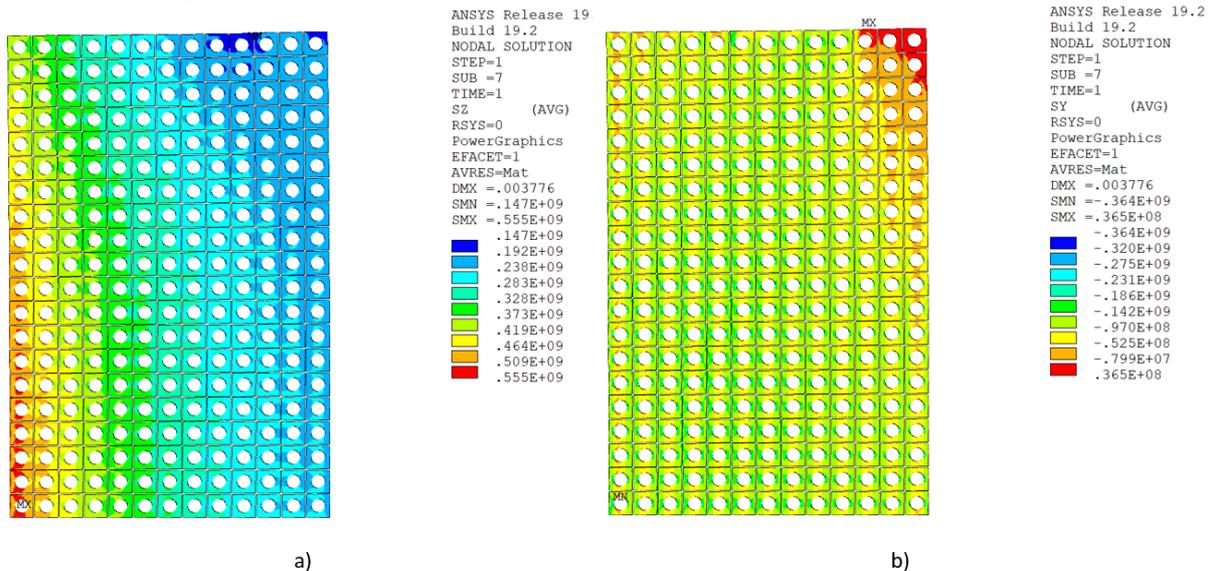

a)        b)

Figure 13. Stress distribution [Pa] in the jackets of coil C3: (left) Hoop stress; (right) Axial stress. Only the stress components in the jacket are shown for the sake of clarity.

A set of coupled global and local axisymmetric models were developed using ANSYS Parametric Design Language (APDL) to analyze the magnetic field and stress distributions in the coils of the target solenoid with different degree of detail. Results from the global models produce the necessary boundary conditions for the local models. The mechanical interaction between cables and jackets is modelled via frictional contacts (μ=0.2).

The ITER Magnet Structural Design Criteria (MSDC) [18] are followed to prescribe allowable values for the stresses in the conductor steel jackets and in the turn insulation. Given the operating conditions of the Target Solenoid, only static criteria are considered. Fatigue will be addressed in future studies if deemed necessary.

Regarding the jackets, assuming a yield strength of 1000 MPa at cryogenic temperature, primary membrane (Pm) and membrane plus bending (Pm+Pb) stress intensities are limited to 667 MPa and 867 MPa, respectively, as shown in Fig. 14.

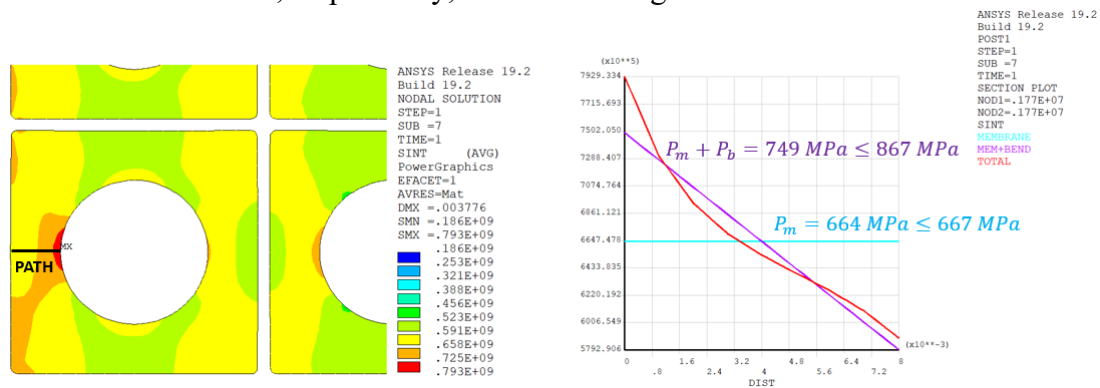

Figure 14. Stress intensity distribution [Pa] at the most loaded jacket and stress linearization along the path defined for the assessment. Only the stress components in the jacket are shown for the sake of clarity.

As for the insulation, in the direction normal to the reinforcing glass fibers the compressive stress must remain below 600 MPa, and tensile strain must be prevented. The maximum allowable shear stress is in the range from 42.5 MPa to 68.6 MPa depending on the local compressive stress. The allowed tensile or compressive strain in the plane of the reinforcing glass fibers is limited to [-0.5%, 0.5%]. The most demanding criterion was found to be the one imposed on the shear stress, which is met everywhere except for a few spots where it is slightly exceeded. These spots are located at jacket corners with low curvature radius, which could be accommodated by a design iteration.

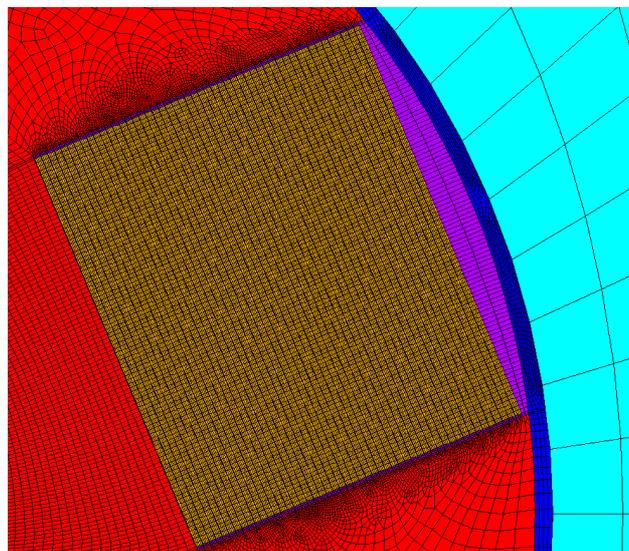

Figure 15. Mesh of the modelled stack of HTS tapes (in orange), solder (in purple), copper former (in red), steel wrap (in dark blue), and jacket (in cyan). Four elements are used along the thickness of each tape.

The ITER MSDC do not cover the mechanical assessment of HTS-based cables. To address this aspect, the focus is put on the tensile and shear stresses in the HTS tapes that lead to debonding and degradation. The cable is modelled in detail for an accurate description of the stress distribution, as depicted in Fig. 15, including the stacked tapes in a soft solder matrix (Young's modulus of 10 GPa) occupying the grooves in the copper former. Smeared orthotropic properties are assigned to the tapes, with main contributions to the stiffness from the Hastelloy and copper layers. Two approaches are adopted to model the interaction between the stack and the former, either the stacks are bonded to the copper, or the stacks are allowed to slide and separate by means of a frictional contact (μ=0.2). In both cases the tapes are bonded by the solder.

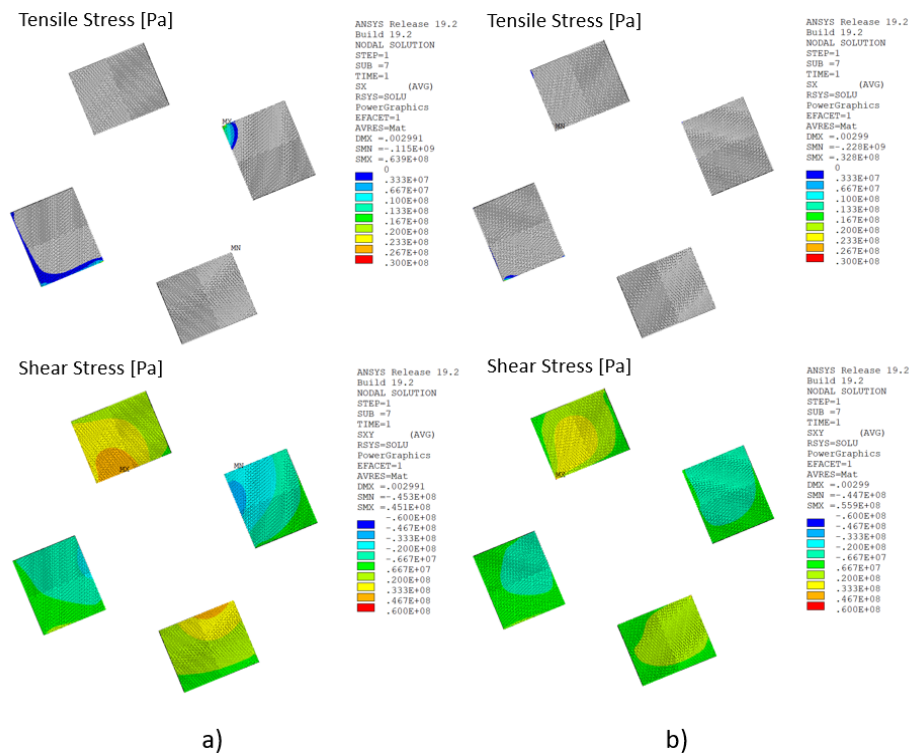

Figure 16 Tensile and shear stress distributions in HTS tapes for: (left) bonded contact between stacks and copper former; (right) frictional contact between stacks and copper former. Grey areas of tensile stress mark areas in compression.

Figure 16 reports the tensile and shear stress distributions in the tapes. The tensile stress is generally below 10 MPa with peak values in some tape corners that can reach up to 60 MPa for the bonded configuration. Allowing the stack to separate from the former substantially mitigates the extent and magnitude of tensile stresses.

The maximum shear stress also decreases with frictional contact from an average around 40 MPa for the stacks bonded to the copper profile, to values in the range of 30 MPa if the stack is not bonded to the copper profile. While detaching the stacks from the grooves seems a satisfactory solution from the tensile stress standpoint, the shear stresses remain considerably high. The presence of such shear stresses is possibly an intrinsic feature of soldered and twisted tape stack, which could explain the degradation observed in short cable samples tested in magnetic field [18, 19]. Solutions aiming at reducing the excessive shear stresses are being explored, looking at the number and width of stacks and their distribution in the copper profile.

**Conclusions and perspective**

We have described the design of a HTS high field and large bore solenoid matching the requirements of the target, decay and capture of a Muon Collider. A double pancake winding using a force-flow cooled superconducting cable largely inspired by the VIPER developed for magnetically confined fusion, seems to be a good solution, meeting most design criteria. Most interesting, it seems possible to achieve high field, 20 T peak field on axis, at high operating temperature, 20 K, which has several benefits. The first is to reduce power consumption, one order of magnitude with respect to the original solution devised in earlier studies. At the same time the design proposed and studied here is more compact, has smaller mass and stored energy compared to previous studies, and has the potential of lower capital and operation expenditure (CAPEX and OPEX). There is clearly much work to do to realize this perceived potential. The study reported here has highlighted some important features and evidenced priority work for the future.

Cooling at high temperature, 20 K, with gaseous helium is not a trivial extrapolation of force-flow supercritical helium near liquid conditions, 4.2 K. High operating pressure, e.g. 20 bar, and larger temperature increase than usual, e.g. 3 K, will be mandatory to avoid excessive distribution losses, and achieving the gain in cryogenic efficiency associated with the higher operating temperature. More studies, integrating the refrigeration cycle, will be necessary to produce an optimal system.

The force-flow cable proposed for the study has already an experimental basis of proven performance [12]. Some additional features have been identified, that could make construction and operation simple. One such example is the reinforcement jacket which has no leak tightness requirement. The studies reported here also show that thermal stability will not be an issue. At the same time quench detection and protection can likely rely on well-established precise voltage measurement, reasonable detection threshold, in the range of 100 mV, and dump voltages within state-of-the-art technology, 5 kV. The hot spot temperature remains well below 200 K in all cases analyzed. It will be very interesting at this point to realize and test samples of the conductor designed here, to confirm manufacturing features, validate the performance reach and margins, and characterize the behavior during quench.

On the side of mechanical design, the overall criteria at the coil level can be satisfied within the allowable limits of common material grades. However, looking at the details of the stress and strain distribution within the cable we may have identified locations and conditions where loads could exceed allowable limits. Tensile and shear stresses at the level of the single tapes could reach values in the range of 60 MPa, whereby it is well known that the internal structure of REBCO tapes is not very resilient to this type of loading, with a wide spread of maximum allowable in the range of a few MPa and up to few tens of MPa. Note that while the analysis was performed for the specific geometry considered here, this may be a result of general applicability to soldered and twisted stacks of tapes. The analysis on this topic is only at the beginning, some avenues have been suggested to resolve this issue, and more optimization work is required, also considering the on-going work in the R&D program being pursued for fusion reactors [19-21]. Also in this case, some strong experimental evidence will be necessary to advance understanding and validate the solutions found.

**Acknowledgements**

This work is partially funded by the European Union under Grant Agreement N. 101094300